\documentclass[journal,12pt,onecolumn,draftclsnofoot,]{IEEEtran}
\IEEEoverridecommandlockouts

\IEEEoverridecommandlockouts

\usepackage{url}
\usepackage{cite}
\usepackage{amsmath,amssymb,amsfonts}
\usepackage{graphbox,graphicx}
\usepackage{textcomp}
\usepackage{xcolor}
\usepackage[capitalize]{cleveref}
\usepackage{algorithmic}
\makeatother   
\usepackage{latexsym,bm}
\usepackage{algorithm}
\usepackage{algorithmic}
\usepackage{setspace}
\usepackage{subcaption}
\usepackage[export]{adjustbox}
\usepackage{float}

\DeclareMathOperator{\Tr}{Tr}

\makeatletter
\def\endthebibliography{%
	\def\@noitemerr{\@latex@warning{Empty `thebibliography' environment}}%
	\endlist
}

\begin{document}
	
	\title{Grassmannian Constellation Design for Noncoherent MIMO Systems Using Autoencoders}
	
	\author{Xiaotian Fu and Didier Le Ruyet, \tiny\thanks{The authors are with CNAM CEDRIC, 292 Rue Saint-Martin, France (Email: xiaotian.fu@lecnam.net; didier.le\_ruyet@cnam.fr)}\normalsize}
	
	
	
	\maketitle
	
	\begin{abstract}
		In this letter, we propose an autoencoder (AE) for designing Grassmannian constellations in noncoherent (NC) multiple-input multiple-output (MIMO) systems. To guarantee the properties of Grassmannian constellations, the proposed AE constructs the transmitted symbols following an unitary space-time modulation. It penalizes the difference between input and output symbols in terms of cross entropy during the training, which is regarded as a generic optimization method. The constellations learned by the proposed AE have substantial symbol error rate (SER) performance gains compared to the non-Grassmannian constellations and conventionally constructed Grassmannian constellations in high SNR regime. The resulting Grassmannian constellation of the proposed AE achieves higher diversity than the non-Grassmannian constellation in i.i.d. Rayleigh channels. Moreover, the proposed approach can be adaptive to different channel statistics by training with corresponding channel realizations.
		
		
	\end{abstract}
	

	\IEEEpeerreviewmaketitle

	\section{Introduction}
	It has been proven that the use of multiple antennas at both transmitter and receiver, also known as multiple-input multiple-output (MIMO) technique, can achieve substantially large spectral efficiency \cite{foschini1998limits}. In a fixed wireless environment, the fading coefficients vary slowly and can be estimated accurately by sending the pilot signals periodically from the transmitter. However, in the mobile environment, the fading coefficients change rapidly, which makes it difficult to accurately estimate the channel state information (CSI) in a limited amount of time. To solve this problem, noncoherent (NC) MIMO systems, where neither the transmitter nor the receiver needs CSI knowledge, have been proposed \cite{grassmann}. The authors in \cite{grassmann} justify the fact that the ergodic capacity of NC MIMO systems at high signal-to-noise ratio (SNR) can be achieved by distributing the transmitted signals isotropically on the (compact) Grassmann manifold, i. e.  every transmitted symbol matrix is unitary. Many conventional approaches have been proposed \cite{4787623,1246045}, however designing optimal Grassmannian constellation is still an open question.
	
	
	After witnessing deep learning (DL) techniques achieving tremendous success in computer vision and natural language processing, more and more applications of DL in communication systems have been proposed \cite{8663966}. Different from other DL applications dedicated to one individual block of communication systems, the autoencoder (AE) comprises transmitter (encoder) neural network, channel layer, and receiver (decoder) neural network which aims at learning the transmitter and receiver jointly for a particular channel \cite{physical_layer}. The transmitter neural network encodes input data into representations, while the receiver neural network reconstructs the input data by decoding the representations corrupted by the channel impairments. Thus, AE technique is known to be an approach to design constellations, as the output of transmitter neural network can be deemed as the transmitted symbols \cite{physical_layer}. Authors in \cite{nc_mimo_ae} adopt AE as a tool for optimizing constellation design in NC MIMO systems. Different from conventional methods designing Grassmannian constellations, the AE in \cite{nc_mimo_ae} constructs non-Grassmannian constellations. The method in \cite{nc_mimo_ae} shows better symbol error rate (SER) performance than conventional approaches when SNR is lower than $20$dB, however it does not follow the theory of NC MIMO systems \cite{grassmann} and it is not rigorously justified to be applicable in NC MIMO systems.
	
	In this letter, a novel AE structure is proposed to design Grassmannian constellations for NC MIMO systems. The main contributions of this letter can be summarized as follows:
	\begin{itemize}
		\item A new approach based on AE techniques to construct Grassmannian constellations is proposed. The proposed AE follows the restriction of Grassmann manifold \cite{grassmann}, where every learned transmitted symbol matrix is unitary. The detailed structure of the proposed AE, including essential orthonormalization process and necessary matrix operations, is well elaborated.
		\item The proposed approach is evaluated by comparing it to various state-of-the-art solutions. Both Grassmannian constellations designed by conventional approaches \cite{vector,4787623,1246045} and the non-Grassmannian constellation \cite{nc_mimo_ae} are studied. The proposed approach significantly outperforms all the state-of-the-art approaches at moderate to high SNRs. Moreover, the resulting constellation of the proposed AE reaches the highest diversity.
		\item SER performance of the different approaches is compared not only in the i.i.d. Rayleigh MIMO channel but also in the correlated fading channel. The proposed AE can be adaptive to the correlated fading channel by training with corresponding channel realizations. Additionally, the constellation for correlated channel learned by the novel approach is efficient for both channels.
	\end{itemize}
	
	The rest of this letter is organized as follows. \cref{sys} introduces the system model and preliminaries of NC MIMO systems. In \cref{pro_ae}, we shed light on the proposed AE. The performance evaluation is in \cref{results}. Finally, \cref{conclusion} concludes the letter.
	
	$Notations:$ Lowercase letters (e.g., \begin{math}x\end{math}) denote scalars, bold lowercase letters (e.g., \begin{math}\mathbf{x}\end{math}) denote column vectors, and bold uppercase letters (e.g., \begin{math}\mathbf{X}\end{math}) denote matrices. $(\cdot)^\dagger$ and $\Tr(\cdot)$ denote the Hermitian transpose and trace operator of matrix. $\mathbf{I}_q$ denotes the $q\times{q}$ identity matrix. Complex Gaussian distribution with mean $\hat{x}$ and variance $\hat{\tau}$ is denoted by $\mathcal{N}_{c}(\hat{x},\hat{\tau})$. 

	\section{System model and preliminaries}
	\label{sys}
	\subsection{System model}
	We consider a NC MIMO communication system where the transmitter and receiver have $N_t$ and $N_r$ antennas, respectively. The channel coefficients are assumed to remain constant for a coherence interval of $T$ and change to a new independent realization in the next time period. We assume that $N_t=\min\{\lfloor{\frac{T}{2}}\rfloor, N_r\}$. $\mathbf{X}\in{\mathbb{C}^{T\times{N_t}}}$ is the transmitted symbol matrix, drawn from the codebook $\mathcal{C}$ with cardinal $|\mathcal{C}|=M$. The received signal is expressed as 
	\begin{equation}
	\mathbf{Y}=\mathbf{X}\mathbf{H}+\sqrt{\frac{N_t}{\rho T}}\mathbf{W}
	\label{re_mat}
	\end{equation}
	where $\mathbf{H}\in{\mathbb{C}^{N_t\times{N_r}}}$ is the channel matrix and $\mathbf{W}\in{\mathbb{C}^{T\times{N_r}}}$ is the additive white Gaussian noise (AWGN) matrix. The entries of $\mathbf{W}$ are drawn independently from $\mathcal{N}_c(0,1)$. $\rho$ denotes system SNR which is independent of the number of transmit antennas $N_t$. To achieve the ergodic capacity of NC MIMO systems, input signals need to represent isotropically distributed $N_t$-dimensional linear subspace of $T$-dimensional complex Euclidean space, $\mathbb{C}^T$ \cite{grassmann}. Accordingly, $\mathbf{X}$ is regarded as a single point on the (compact) Grassmann manifold $\mathbb{G}_{N_t}(\mathbb{C}^T)$. Mathematically, $\mathbf{X}$ is an unitary matrix, following the constraint
	\begin{equation}
	\mathbf{X}^\dagger\mathbf{X}=\mathbf{I}_{N_t}
	\label{unitary_com}
	\end{equation}
	
	
	
	
	Channel of the system model is written as 
	\begin{equation}
	\mathbf{H}=(\mathbf{R}^t)^{1/2}\mathbf{G}(\mathbf{R}^r)^{1/2}
	\end{equation}
	where the elements of matrix $\mathbf{G}\in\mathbb{C}^{N_t\times{N_r}}$ are i.i.d. and drawn from $\mathcal{N}_c(0,1)$, and $\mathbf{R}^t$ and $\mathbf{R}^r$ are called the transmit and receive covariance matrix, respectively. We consider two types of transmission channel, i.i.d. Rayleigh and correlated fading channel. For the correlated fading channel, we assume that only the receive antennas are correlated and the receive covariance matrix is modelled using the exponential model. Therefore, the transmit covariance matrix is $\mathbf{R}^t=\mathbf{I}_{N_t}$ and the entries of the receive covariance matrix are defined as
	\begin{equation}
	R^r_{i,j}=\{r^{|i-j|}\},\quad{r\in[0,1)}
	\end{equation}
	where $i,j\in\{1,2\cdots,N_r\}$ are the row and column indexes of matrix $\mathbf{R}^r$, respectively.
	
	Considering the NC transmission scenario and the unitary matrix $\mathbf{X}$, the maximum-likelihood (ML) detection at the receiver is given as \cite{ust_mod}
	\begin{equation}
	\begin{aligned}
	\hat{\mathbf{X}}&=\arg\max_{\mathbf{X}}p(\mathbf{Y}|\mathbf{X})\\
	&=\arg\max_{\mathbf{X}}\Tr\{\mathbf{Y}^\dagger\mathbf{X}\mathbf{X}^\dagger\mathbf{Y}\}
	\end{aligned}
	\label{detection}
	\end{equation}
	
	\subsection{Grassmannian constellations}
	
	
	The constellation generation by a classical method can be regarded as a problem of maximization of a  distance metric which implies defining an appropriate metric to measure the distance between the constellation points. Following the perturbation analysis of the received signal given in \cite{4787623}, chordal Frobenius distance (norm) is the appropriate metric for the Grassmannian constellation design. The chordal Frobenius distance between unitary matrix $\mathbf{X}_1\in\mathcal{C}$ and $\mathbf{X}_2\in\mathcal{C}, \mathbf{X}_2\neq\mathbf{X}_1$ is expressed as
	\begin{equation}
	d(\mathbf{X}_1, \mathbf{X}_2)=\sqrt{2N_t-2\Tr(\bm{\Sigma}_{\mathbf{X}_1,\mathbf{X}_2})}
	\label{dist}
	\end{equation}
	where $\bm{\Sigma}_{\mathbf{X}_1,\mathbf{X}_2}$ is the diagonal matrix containing the singular values of $\mathbf{X}_1^\dagger\mathbf{X}_2$. 
	

	\section{Proposed autoencoder for Grassmannian constellation design}
	\label{pro_ae}
	To facilitate the application of neural networks, the input-output equation given in  \cref{re_mat} is rewritten using the  equivalent real-valued notation:
	\begin{equation}
	\bar{\mathbf{Y}}=\bar{\mathbf{X}}\bar{\mathbf{H}}+\sqrt{\frac{N_t}{\rho T}}\bar{\mathbf{W}}
	\end{equation}
	where 
	\begin{eqnarray*}
		\begin{aligned}
			\bar{\mathbf{X}}&=\begin{bmatrix}
				\Re\{\mathbf{X}\} & -\Im\{\mathbf{X}\}\\
				\Im\{\mathbf{X}\} & \Re\{\mathbf{X}\}
			\end{bmatrix}\quad
			\bar{\mathbf{H}}=\begin{bmatrix}
				\Re\{\mathbf{H}\} & -\Im\{\mathbf{H}\}\\
				\Im\{\mathbf{H}\} & \Re\{\mathbf{H}\}
			\end{bmatrix}\\
			\bar{\mathbf{Y}}&=\begin{bmatrix}
				\Re\{\mathbf{Y}\} & -\Im\{\mathbf{Y}\}\\
				\Im\{\mathbf{Y}\} & \Re\{\mathbf{Y}\}
			\end{bmatrix}\quad
			\bar{\mathbf{W}}=\begin{bmatrix}
				\Re\{\mathbf{W}\} & -\Im\{\mathbf{W}\}\\
				\Im\{\mathbf{W}\} & \Re\{\mathbf{W}\}
			\end{bmatrix}
		\end{aligned}
	\end{eqnarray*}
	$\mathbf{X}$, $\mathbf{H}$, $\mathbf{Y}$ and $\mathbf{W}$ are the complex-valued matrices used in \cref{re_mat}. Consequently, \cref{unitary_com} is rewritten as
	\begin{equation}
	\bar{\mathbf{X}}^\dagger\bar{\mathbf{X}}=\mathbf{I}_{2N_t}
	\label{ortho_real}
	\end{equation}
	
	\label{proposed_ae}
	\begin{figure*}[t]
		\centering
		\centerline{\includegraphics[width=16cm,keepaspectratio]{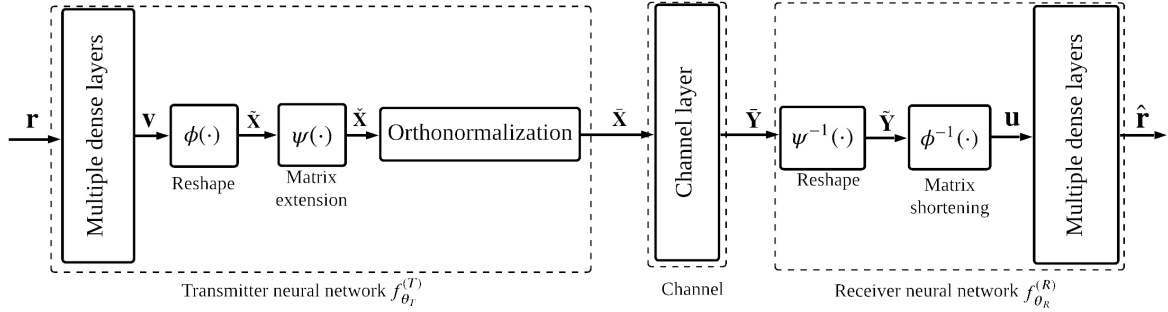}}
		\caption{Structure of the proposed AE in NC MIMO}
		\label{proposed_ae_strucutre}
	\end{figure*}
	
	\cref{proposed_ae_strucutre} illustrates the detailed structure of the proposed AE. Functions $f^{(T)}_{\bm{\theta}_T}(\cdot)$ and $f^{(R)}_{\bm{\theta}_R}(\cdot)$ are used to represent the transmitter and receiver neural network, respectively. Because $\bar{\mathbf{X}}$ and $\bar{\mathbf{Y}}$ are in a specific matrix format, both $f^{(T)}_{\bm{\theta}_T}(\cdot)$ and $f^{(R)}_{\bm{\theta}_R}(\cdot)$ contain not only multiple neural layers but also matrix operation blocks. $\bm{\theta}_T$ and $\bm{\theta}_R$ are the weight and bias set of multiple neural layers inside $f^{(T)}_{\bm{\theta}_T}(\cdot)$ and $f^{(R)}_{\bm{\theta}_R}(\cdot)$, respectively. The channel layer is built by a single neural layer where weights represent channel coefficients and biases represent additional noise. Stress that the channel layer is not trainable. The multiple neural layers at both sides are assumed to be fully-connected. Vector $\mathbf{O}_T=[O_{T,0}, O_{T,1},\cdots,O_{T,l_T+1}]$ and $\mathbf{O}_R=[O_{R,0}, O_{R,1},\cdots,O_{R,l_R+1}]$ denote the neuron number of each fully-connected layer at the transmitter and receiver, respectively. $l_T$ is the number of hidden layers in $f^{(T)}_{\bm{\theta}_T}$ and $l_R$ is the number of hidden layers in $f^{(R)}_{\bm{\theta}_R}$. 
	
	The input of the proposed AE is one-hot vector $\mathbf{r}$ with $M\times{1}$ binary bits, in which only one element, indicating the transmitted message, is one and the others are zero. Notice that $r_n=\begin{cases}1,\quad n=m\\
		0,\quad\text{otherwise}
		\end{cases}(1\leqslant{n}\leqslant{M})$, where $r_n$ is the $n$-th element in $\mathbf{r}$ and $m$ is the transmitted message.
	The input vector is passed through the first multiple dense layers, then matrix operations and finally orthonormalization process. Vector $\mathbf{v}\in\mathbb{R}^{2TN_t\times1}$ is the output of the multiple dense layers at the transmitter with input $\mathbf{r}$. Since the transmitted signal is a matrix, it is necessary to tailor vector $\mathbf{v}$. Firstly, reshaping function $\bm{\phi}(\cdot):\mathbb{R}^{2TN_t\times{1}}\rightarrow\mathbb{R}^{2T\times{N_t}}$ converts column vector $\mathbf{v}$ into matrix $\tilde{\mathbf{X}}$, formulated as
	\begin{equation}
	\tilde{\mathbf{X}}=\bm{\phi}(\mathbf{v})=\begin{bmatrix}
	\tilde{\mathbf{X}}_1\\
	\tilde{\mathbf{X}}_2
	\end{bmatrix}
	\end{equation}
	where $\tilde{\mathbf{X}}_1\in\mathbb{R}^{T\times{N_t}}$ and $\tilde{\mathbf{X}}_2\in\mathbb{R}^{T\times{N_t}}$ are upper and lower half of $\tilde{\mathbf{X}}$, respectively. Then, function $\bm{\psi}(\cdot):\mathbb{R}^{2T\times{N_t}}\rightarrow\mathbb{R}^{2T\times{2N_t}}$ enlarges the matrix $\tilde{\mathbf{X}}$ to $\check{\mathbf{X}}$ by duplicating the entries in a particular way, which is expressed as 
	\begin{equation}
	\check{\mathbf{X}}=\bm{\psi}(\tilde{\mathbf{X}})=\bm{\psi}\bigg(\begin{bmatrix}
	\tilde{\mathbf{X}}_1\\
	\tilde{\mathbf{X}}_2
	\end{bmatrix}\bigg)
	=\begin{bmatrix}
	\tilde{\mathbf{X}}_1 & -\tilde{\mathbf{X}}_2\\
	\tilde{\mathbf{X}}_2 & \tilde{\mathbf{X}}_1
	\end{bmatrix}
	\end{equation}
	
	To satisfy the requirement of Grassmannian constellations, orthonormalization is adopted to ensure that the output matrices of $f^{(T)}_{\bm{\theta}_T}$ are unitary as demonstrated in \cref{ortho_real}. Two methods can be applied to carry out the orthonormalization. 
	\begin{itemize}
		\item{Method 1.} 
		This is based on the square root of $\check{\mathbf{X}}^\intercal\check{\mathbf{X}}$ which is written as 
		\begin{equation}
		\bar{\mathbf{X}}=\check{\mathbf{X}}(\check{\mathbf{X}}^\intercal\check{\mathbf{X}})^{-1/2}
		\label{eq_ortho}
		\end{equation} 
		\item{Method 2.} If the  matrix $\check{\mathbf{X}}$ can be decomposed by applying the singular value decomposition (SVD), it is expressed as $\check{\mathbf{X}}=\mathbf{U}\mathbf{\Sigma}\mathbf{V}^\intercal$. Therefore, \cref{eq_ortho} can be rewritten as
		\begin{equation}
		\bar{\mathbf{X}}=\mathbf{U}\mathbf{\Sigma}\mathbf{V}^\intercal\big(\mathbf{V}\mathbf{\Sigma}^{-1}\mathbf{V}^\intercal\big)=\mathbf{U}\mathbf{V}^\intercal
		\label{eq_ortho_new}
		\end{equation} 
	\end{itemize}
	
	%
	
	In the receiver neural network, $\bm{\phi}^{-1}(\cdot)$ and $\bm{\psi}^{-1}(\cdot)$ are the inverse functions of $\bm{\phi}(\cdot)$ and $\bm{\psi}(\cdot)$, respectively. They are utilized to convert matrices to column vectors. The input of the multiple dense layers within $f^{(R)}_{\bm{\theta}_R}$ is $\mathbf{u}\in\mathbb{R}^{2TN_r\times{1}}$, which is obtained by
	\begin{equation}
	\mathbf{u}=\bm{\phi}^{-1}(\tilde{\mathbf{Y}})=\bm{\phi}^{-1}\big(\bm{\psi}^{-1}(\bar{\mathbf{Y}})\big)
	\end{equation}
	The output of multiple dense layers in $f^{(R)}_{\bm{\theta}_R}(\cdot)$ is vector $\hat{\mathbf{r}}\in[0,1]^{M\times{1}}$ following $||\hat{\mathbf{r}}||_1=1$, which contains the probabilities of all corresponding messages. Hence, the $Softmax$ activation function is applied at the last dense layer at the receiver. In the proposed AE, loss function is defined as categorical cross entropy to penalize the difference between $\mathbf{r}$ and $\hat{\mathbf{r}}$. Therefore, it is defined as
	\begin{equation}
	\mathcal{L} = -\sum_{i=1}^{S}\|diag(\mathbf{r}^{(i)})\log(\hat{\mathbf{r}}^{(i)})\|_1
	\end{equation}
	where $\log(\cdot)$ means the element-wise logarithm operation of vectors and $S$ denotes the training batch size. Please note that instead of maximizing the metric in \cref{dist}, the proposed AE optimizes the Grassmannian constellations by minimizing the loss function values.

	\section{Numerical results}
	\label{results}
	In this section, we evaluate the proposed AE by comparing with four different state-of-the-art approaches. For fair comparison, both conventional approaches including Lloyd algorithm \cite{vector}, greedy algorithm \cite{4787623} and $G_{4,2}$ constellation \cite{1246045} and the AE \cite{nc_mimo_ae} constructing non-Grassmannian constellations are studied. In the simulation, we build neural networks on the DL framework Pytorch \cite{pytorch}. We assume the number of transmit antennas $N_t=2$, number of received antennas $N_r=2$, coherence interval $T=4$ and constellation size $M=256$.
	

	
	\begin{table} [t]
		\caption{Structure and training hyper-parameters of the proposed AEs in NC MIMO}
		\vspace{-3mm}
		\begin{center}
			\begin{tabular}{c|c}%
				\hline
				Name & Proposed AE1/ AE2 \\ \hline
				$M$ & $256$  \\ \hline
				$\mathbf{O}_T$  & $[M,20M,20M,2TN_t]$  \\ \hline
				$\mathbf{O}_R$ & $[2TN_r,20M,20M,M]$  \\ \hline
				$S$ & $2000$  \\ \hline
				Nb of epochs &  $50$  \\ \hline
				Nb of batches &  $50$  \\ \hline
				Learning rate & $0.0003$  \\ \hline
				Optimizer & Adam   \\ \hline
				SNR & $15 dB$  \\ \hline
				Channel & i.i.d. Rayleigh / Correlated  $r=0.9$  \\ \hline
			\end{tabular}
			\label{paras_nc_mimo}
		\end{center}
	\end{table}
	To investigate performance of the proposed AE over different channels, we trained it with the same structure and hyper-parameters but different channel layers. We differentiate two training by calling the one trained in i.i.d. Rayleigh channel proposed AE1 and the one trained in correlated fading channel proposed AE2. \cref{paras_nc_mimo} gives structure and details of these two AE training. In the proposed AE, the rectified linear units (ReLU) activation function is applied at the intermediate layers and Xavier method \cite{weight} is applied for the weights initialization. We implement the orthonormalization with method 1 due to the back-propagation constraint of the SVD operation in Pytorch. It is worth stressing that AEs are optimization methods for constellation design and are only involved in the training stage. After the offline-training, the learned constellations are implemented into look-up tables at the transmitter and the ML detection \cref{detection} is adopted at the receiver. 
	
	\begin{figure}[t]
		\centering
		\centerline{\includegraphics[width=12cm,keepaspectratio]{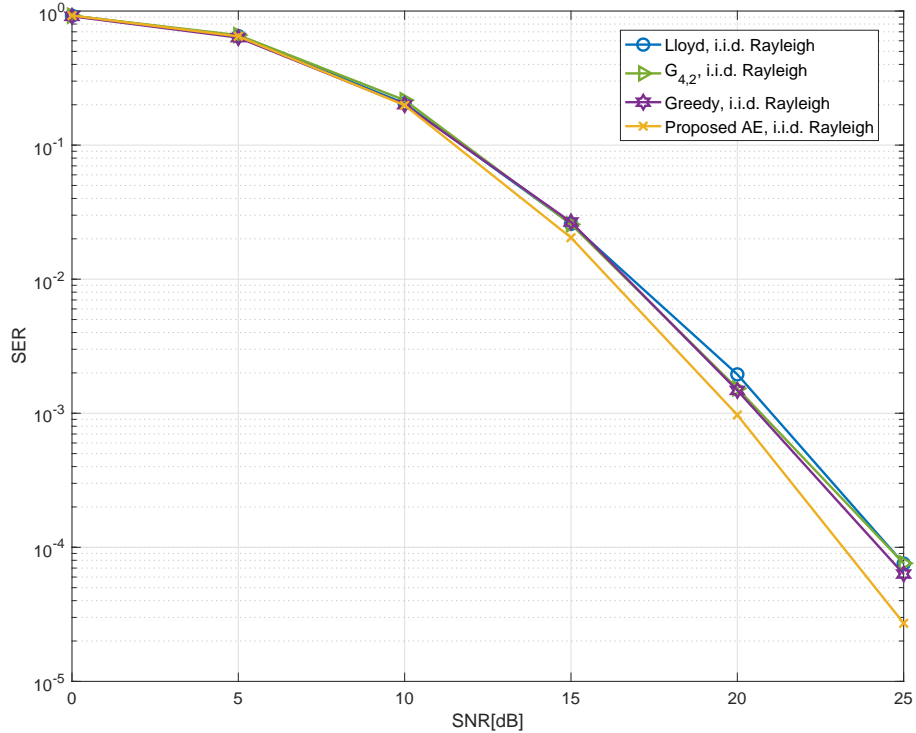}}
		\caption{SER comparison of different constellations in i.i.d. Rayleigh channel}
		\label{ser_nc_mimo_rayleigh}
	\end{figure}
	
	\cref{ser_nc_mimo_rayleigh} compares the SER performance of different constellations in i.i.d. Rayleigh channel. The label AE\ reference refers to the non-Grassmannian constellation learned by the AE in \cite{nc_mimo_ae}. Because of constructing non-unitary space-time constellations, the AE reference merely has better performance than Grassmannian constellations at the SER level of $10^{-2}$. Nevertheless, the proposed AE1 starts to outperform the AE reference from $SNR=17$dB, while other Grassmannian constellations start to outperform it from $SNR\simeq20$dB. The constellation proposed AE1 outperforms Grassmannian counterparts when SNR is greater than $10$dB. At $SNR=20$dB, it outperforms the greedy approach and $G_{4,2}$ constellations by $0.7$dB, Lloyd approach by $1.1$ dB and the non-Grassmannian constellation  by $1.4$dB. When generating Grassmannian constellations, the novel approach considers the  cross entropy metric between the input and the output, rather than chordal Frobenius distance used by the conventional approaches. Therefore, intuitively, the newly proposed approach is more efficient to design Grassmannian constellations regarding lower SER. Additionally, the SER curves of Grassmannian constellations are rather parallel at high SNRs, while the slope of the non-Grassmannian constellation curve is more gentle than the former. We can see that the constellation proposed AE1 can reach diversity of almost $3$, whereas the non-Grassmannian constellation has diversity of $2$.
	
	\begin{figure}[t]
		\centering
		\centerline{\includegraphics[width=12cm,keepaspectratio]{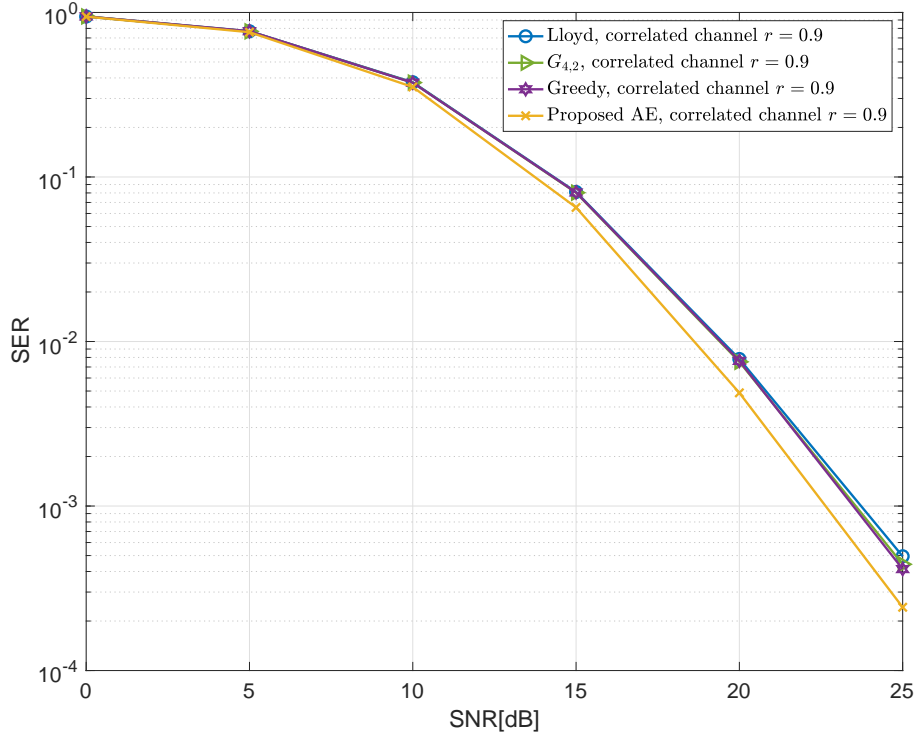}}
		\caption{SER comparison of different constellations in correlated fading channel with $r=0.9$}
		\label{ser_nc_mimo_correlated}
	\end{figure}
	
	\cref{ser_nc_mimo_correlated} illustrates the SER performance comparison of different constellations in correlated fading channel with $r=0.9$. As shown, the correlated-AE reference and the AE reference have almost the same SER performance in correlated fading channels. It implies that considering the correlation model does not bring any SER performance enhancement for non-Grassmannian constellations. However, the constellation proposed AE2 outperforms the constellation proposed AE1 by $1$dB at $SNR=20$dB over correlated fading channel, since the former is trained with the corresponding channel realizations. This shows that the proposed approach can be adaptive to different channel conditions by considering the channel realizations in the training.
	
	\begin{figure}[t]
		\centering
		\centerline{\includegraphics[width=12cm,keepaspectratio]{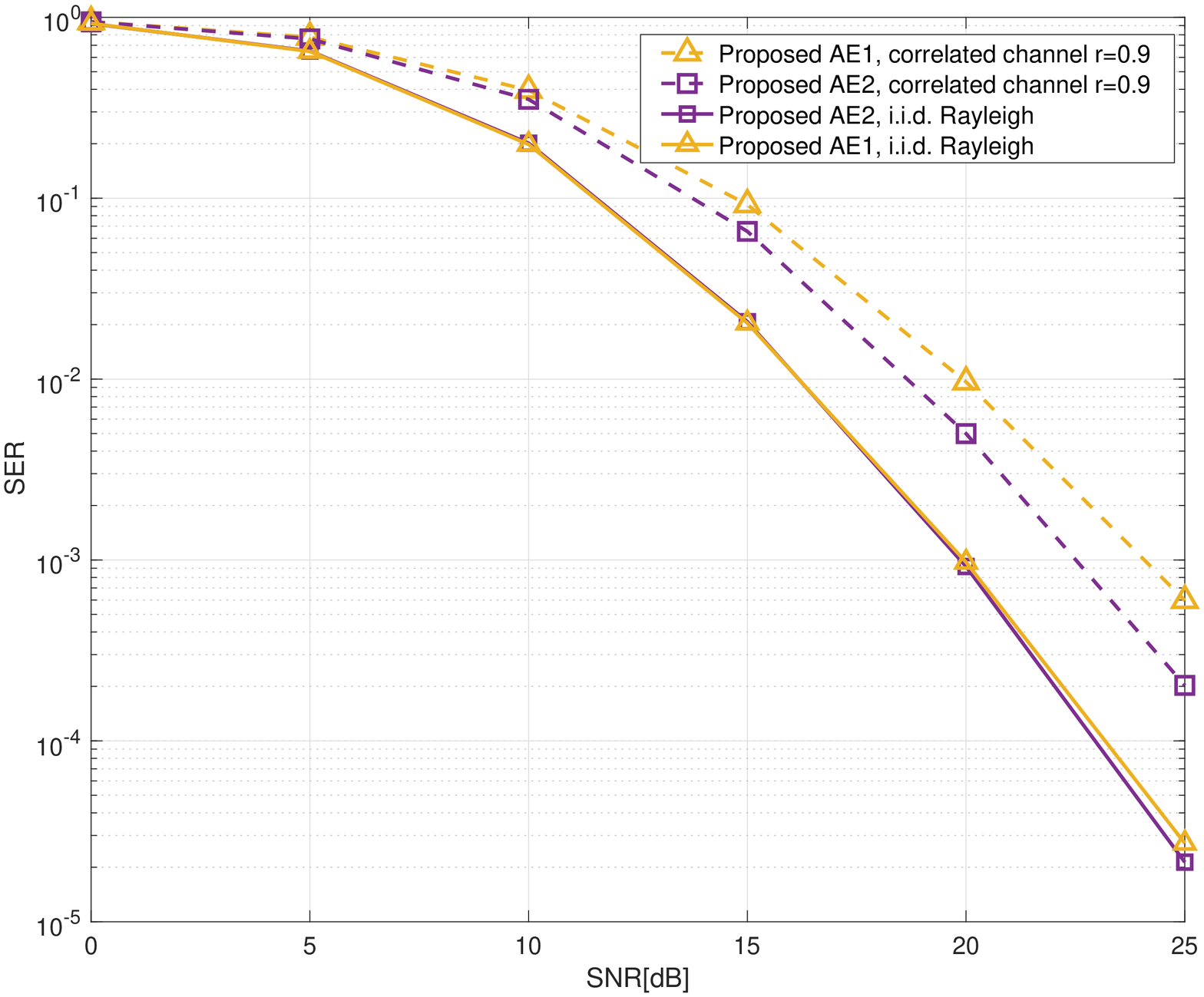}}
		\caption{SER comparison of constellations learned by the novel approach over different channels}
		\label{ser_nc_mimo_AE}
	\end{figure}
	
	\cref{ser_nc_mimo_AE} compares the SER performance of constellations learned by the novel approach over different channels. The constellations constructed for different channels are investigated in i.i.d. Rayleigh and correlated fading channels. The constellation proposed AE2 significantly outperforms the proposed AE1 in the correlated fading channel. Meanwhile, it can achieve similar SER performance as the proposed AE1 in i.i.d. Rayleigh channel. Thus, the proposed AE2 is a universal solution for both i.i.d. Rayleigh and correlated fading channel.
	
\begin{figure}[t]
		\minipage{0.5\textwidth}
		\centering\includegraphics[width=7cm,keepaspectratio]{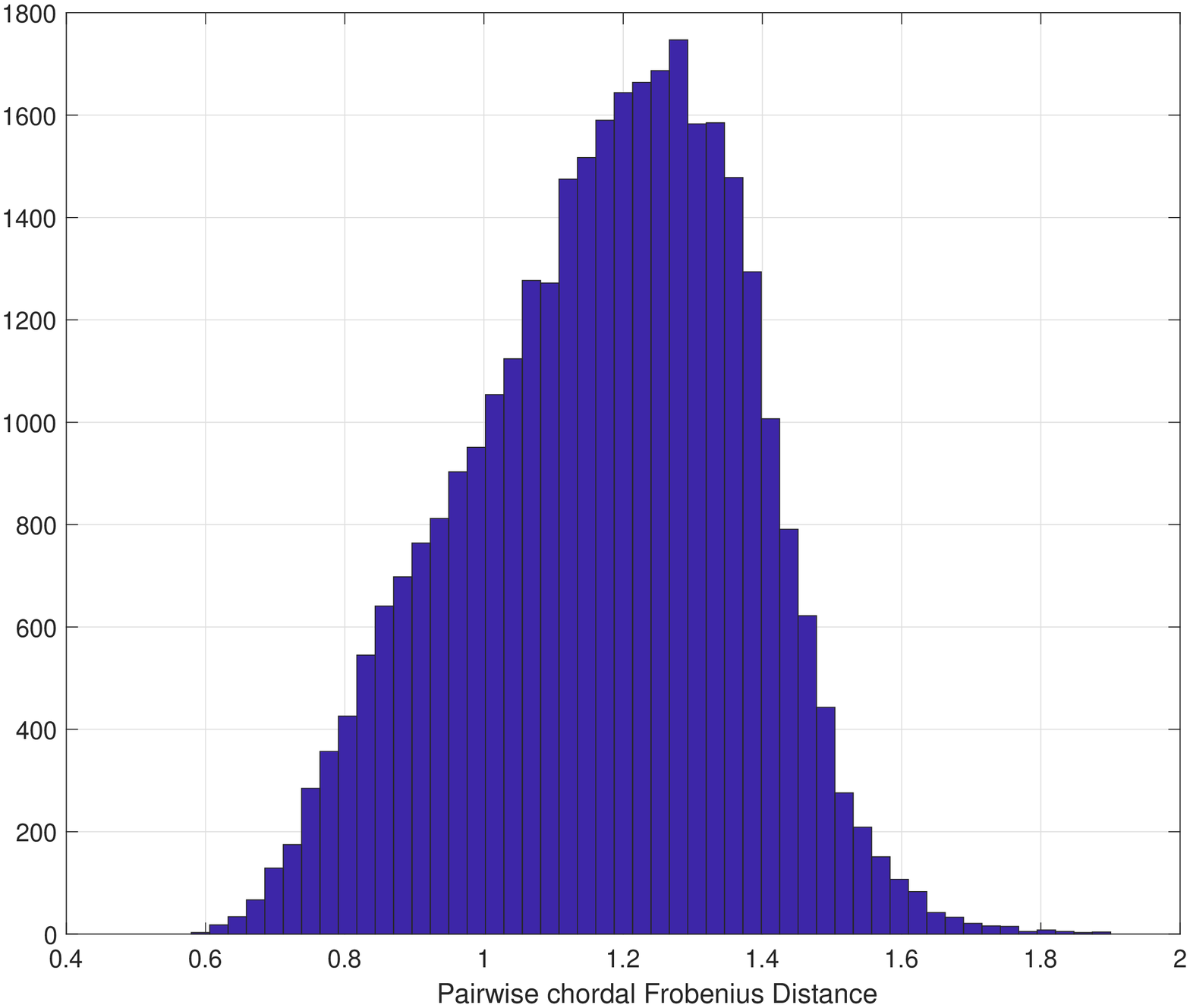}
		\subcaption{Constellation by $Lloyd$ ($d_{mean}=1.1731$)}
		\label{cf_frass}
		\endminipage\hfill
		\minipage{0.5\textwidth}
		\centering\includegraphics[width=7cm,keepaspectratio]{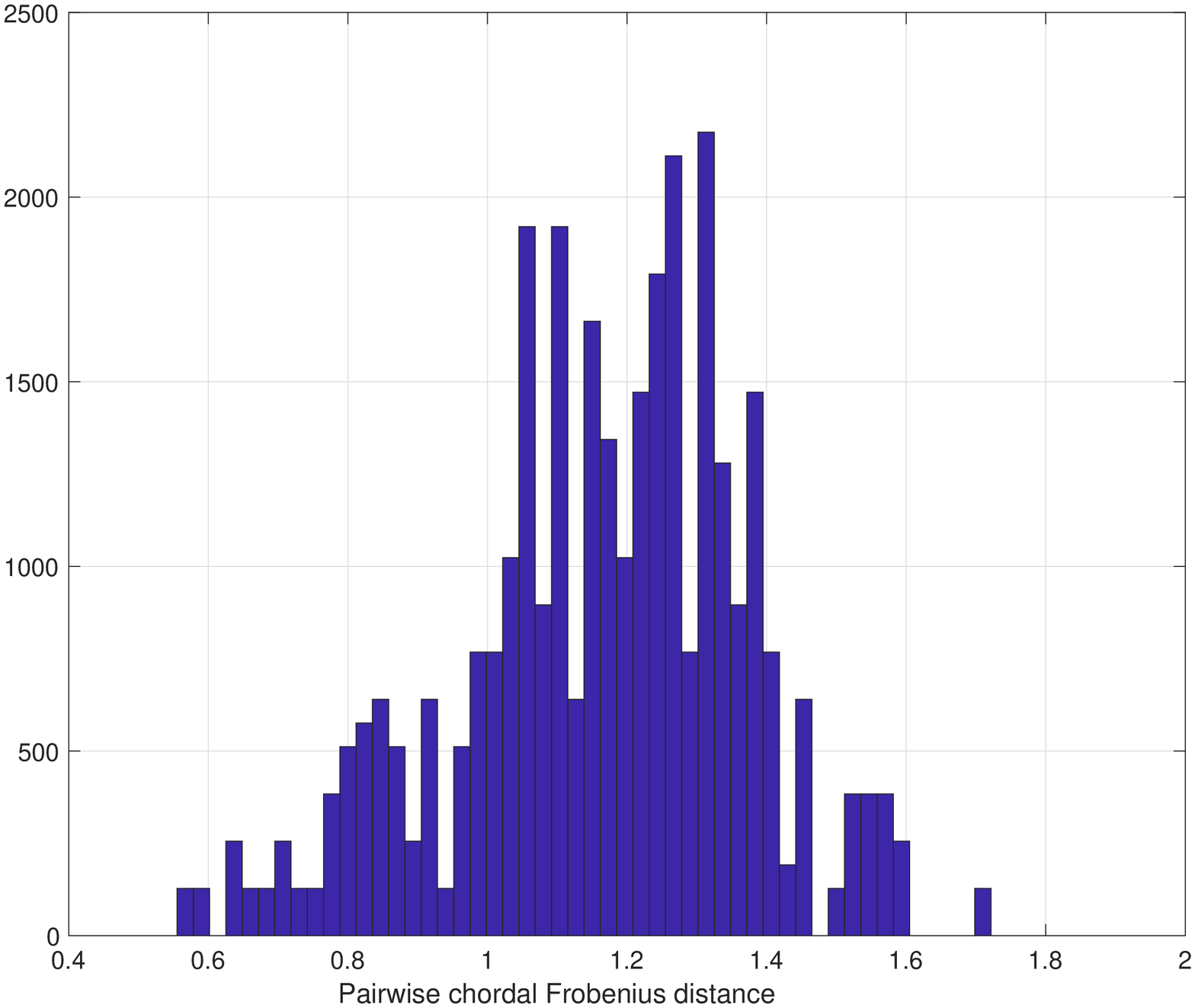}
		\subcaption{Constellation $G_{4,2}$ \cite{1246045} ($d_{mean}=1.1634$)}
		\label{cf_C1}
		\endminipage\hfill\\
		\\
		\minipage{0.5\textwidth}
		\centering\includegraphics[width=7cm,keepaspectratio]{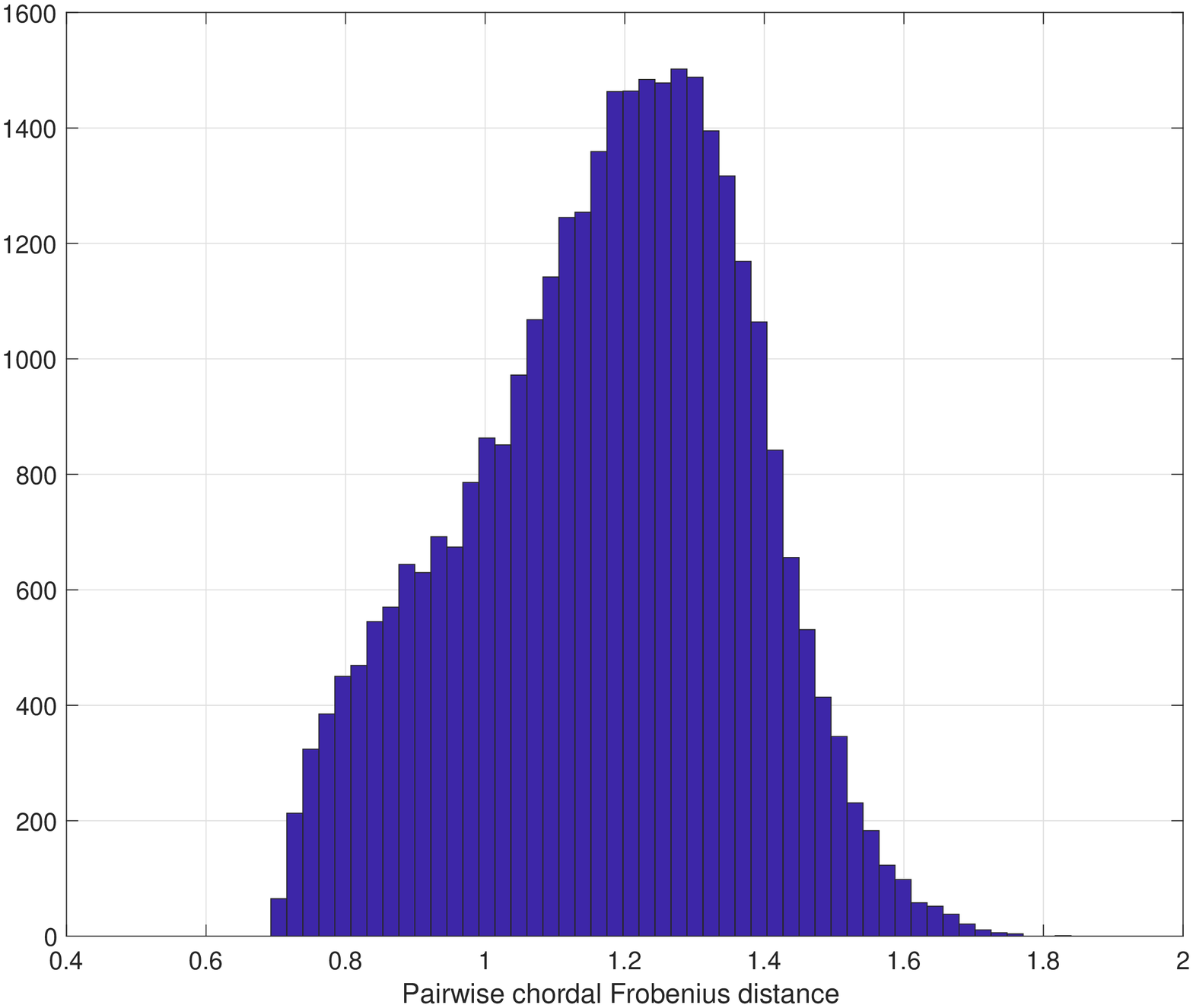}
		\subcaption{Constellation by greedy approach \cite{4787623} $(d_{mean}=1.1730)$}
		\label{cf_ae_correlated}
		\endminipage\hfill
		\minipage{0.5\textwidth}
		\centering\includegraphics[width=7.5cm,keepaspectratio]{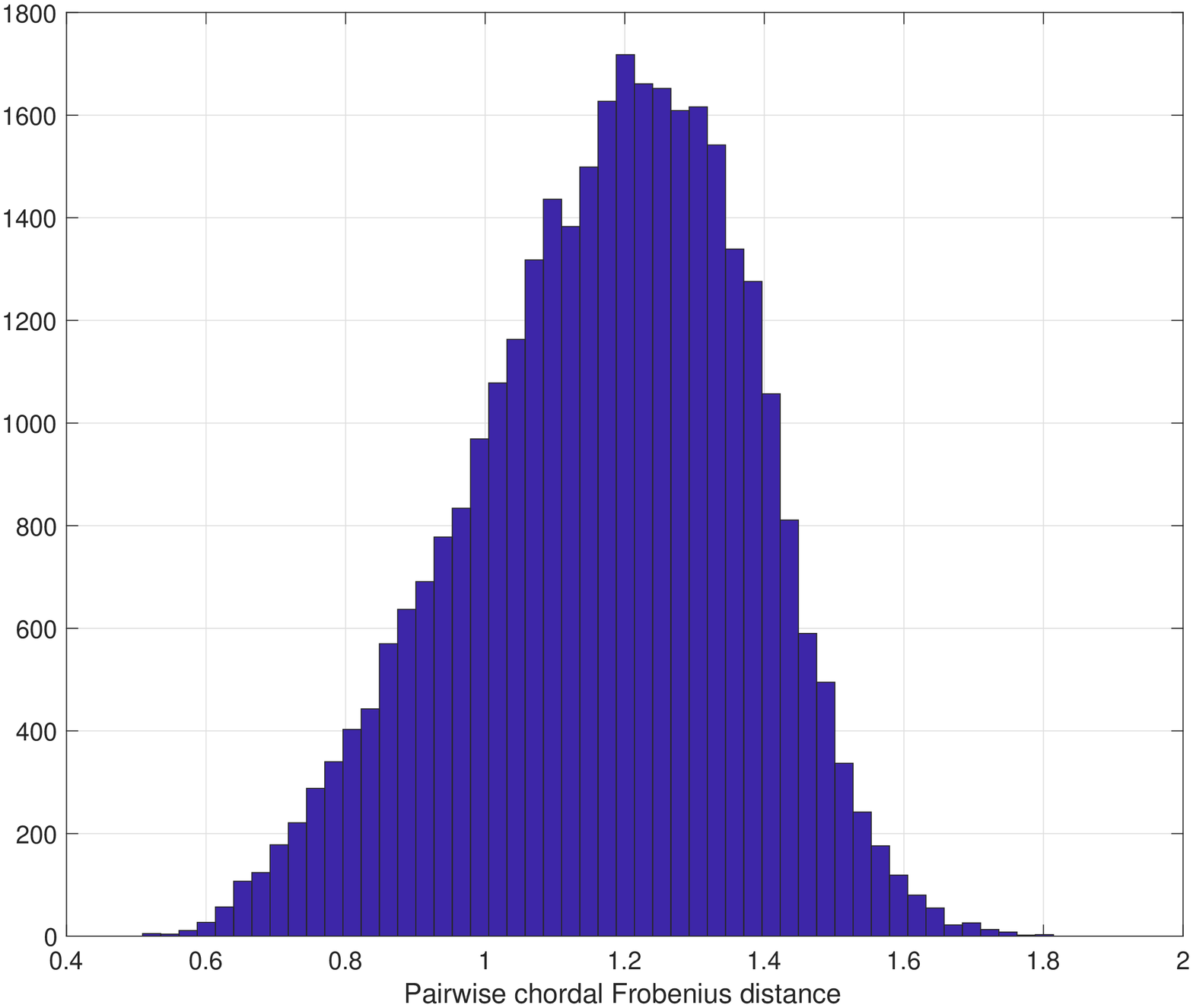}
		\subcaption{Constellation by the proposed AE1 $(d_{mean}=1.1727)$}
		\label{cf_ae_rayleigh}
		\endminipage\hfill 
		\caption{Distribution of pairwise chordal Frobenius distance between constellation points}
		\label{cf}
	\end{figure}

	\cref{cf} shows the distribution and mean of the pairwise chordal Frobenius distance between the Grassmannian constellation points. It is notable that the distribution of constellation $G_{4,2}$ is different from the others since it is algebraically designed for the system with $T=4$ and $N_t=2$ while the others are generic optimization methods for NC MIMO systems. The proposed AE1 has a negligible mean distance gap compared to the other generic approaches, Lloyd and greedy.
	
	
	\section{Conclusion}
	\label{conclusion}
	We propose a novel optimization approach based on AE technique to design Grassmannian constellations for NC MIMO systems. To satisfy the requirement of Grassmannian constellations, orthonormalization and essential matrix operations are included in the proposed AE. Four state-of-the-art solutions are compared with the proposed approach in the performance evaluation. The constellations learned by the novel approach significantly outperforms the conventionally designed constellations and non-Grassmannian constellations in moderate and high SNR regime. It also allows the system to have larger diversity than other constellations. Moreover, it can be adaptive to different channel statistics by training with corresponding channel realizations. It might be suggested that the proposed AE is a promising optimization method of Grassmannian constellation design for all SNRs in diverse channel conditions.
	
	\bibliographystyle{IEEEtran}
	\bibliography{bib_ae_nc_mimo}
	
\end{document}